\documentclass[a4paper,fleqn,usenatbib]{mnras}

\usepackage[T1]{fontenc}
\usepackage{ae,aecompl}

\usepackage{graphicx,epsfig,txfonts}	

\title[Short title, max. 45 characters]{Detection of a cyclotron line in SXP 15.3 during its 2017 outburst}

\author[C. Maitra et al.]{
C.~Maitra,$^{1}$\thanks{E-mail: cmaitra@mpe.mpg.de}
B.~Paul,$^{2}$ 
F.~Haberl,$^{1}$
G.~Vasilopoulos$^{1}$
\\
$^{1}$Max-Planck-Institut f{\"u}r extraterrestrische Physik, Giessenbachstra{\ss}e 1, 85748 Garching, Germany \\
$^{2}$Raman Research Institute, C.V. Raman Avenue, Sadashivanagar, Bangalore 560064, India}

\date{Accepted XXX. Received YYY; in original form ZZZ}

\pubyear{2018}

\newcommand{\astrosat}{{\it AstroSat}}
\newcommand{\nus}{{\it NuSTAR}}
\newcommand{\swi}{{\it Swift}}
\newcommand{\cha}{{\it Chandra}}
\newcommand{\sxp}{SXP\,15.3}

\newcommand{\ergcm}[1]{erg cm$^{-2}$ s$^{-1}$}
\newcommand{\ergs}[1]{erg s$^{-1}$}

\begin{document}
\label{firstpage}
\pagerange{\pageref{firstpage}--\pageref{lastpage}}
\maketitle

\begin{abstract}
We report the results of \astrosat\ and \nus\ observations of the Be/X-ray binary pulsar SXP 15.3 in the Small Magellanic Cloud 
during its outburst in late 2017, when the source reached a luminosity level of $\sim$10$^{38}$ erg\,s$^{-1}$, close to the Eddington limit. 
The unprecedented broadband coverage of the source allowed us to perform timing and spectral analysis between 3 and 80 keV. 
The pulse profile exhibits a significant energy dependence, and morphs from a double peaked profile to a single broad pulse at energies $>15$\,keV.
This can be explained by a spectral hardening during an intensity dip seen between the two peaks of the pulse profile.
We detect a Cyclotron Resonance Scattering Feature (CRSF) at $\sim$5 keV in the X-ray spectrum, independent of the choice of 
the continuum model. This indicates a magnetic field strength of $6\times10^{11}$ G for the neutron star.

\end{abstract}




\section{Introduction}
\label{sec:intro}
\sxp\ (aka RX J0052.1--7319) is a transient X-ray binary pulsar located in the Small Magellanic Cloud (SMC), first detected using 
{\it Einstein} observations \citep{1992ApJS...78..391W}.
Later the source was also detected in the {\it ROSAT}-PSPC data as a hard and highly variable source and classified as a transient X-ray binary 
candidate \citep{1996A&A...312..919K}. Coherent pulsations with a period of 15.3 s were discovered in 1996 using {\it ROSAT} and {\it CGRO} observations with a pulse fraction of 27\% at a 
luminosity (0.1--2 keV) of $\sim$10$^{37}$ erg\,s$^{-1}$ \citep{1999IAUC.7081....4L,2001ApJ...560..378F}. Subsequently \cite{2000A&A...354..999K} 
investigated the {\it ROSAT}-HRI observations in 1995 and 1996 and found a large variation 
in the flux by a factor of $\sim$200, further ascertaining its transient nature. The optical counterpart to the source was identified as a likely Be star by \cite{1999IAUC.7101....1I}, which was later confirmed
as an O9.5IIIe star \citep[$V=14.6$ mag,][]{2001A&A...374.1009C}. The source has not been reported in an outburst or a bright state ever since until July 25 2017, when the {\it Swift}
SMC Survey (S-CUBED) detected a brightening of the source \citep{2017ATel10600....1K}. Pulsations at 15.253 s were detected, and the absorption corrected luminosity (0.5--10 keV) corresponded to  $2.4\times10^{37}$ erg\,s$^{-1}$.
The optical counterpart also exhibited a corresponding brightening. The source re-brightened again in November 2017, reaching a higher X-ray luminosity of $3.9\times10^{37}$ erg\,s$^{-1}$
\citep{2017ATel11030....1D}. This triggered several Target of Opportunity observations (ToO).

In this letter, we present the broadband timing and spectral characteristics of \sxp\ for the first time, using \astrosat\ and \nus\ observations performed
during the recent outburst which started in October 2017. In Sect. 2 we describe the observations and data reduction. 
We present the results of a timing analysis (Sect. 3), a spectral analysis (Sect. 4) and pulse phase-resolved spectroscopy (Sect. 5). 
Discussions and conclusions are presented in Sect. 6. 
\section{Observations and data reduction}
Following the report of an outburst in November 2017, \sxp\ was observed with \nus\ \citep{2013ApJ...770..103H} on 2017--11--30 for $\sim$70\,ks as a 
ToO observation. A simultaneous \swi/XRT \citep{2005SSRv..120..165B} observation was also carried out for 3\,ks (Obsid 00088639001).  
In addition, we triggered a ToO observation of the source with \astrosat\ \citep{2006AdSpR..38.2989A,2014SPIE.9144E..1SS}. 
The observation was performed on 2017--12--08 with an exposure of 60\,ks. The simultaneous \swi\ and \nus\ observation will be referred hereafter
as Obs. 1 and the \astrosat\ observation as Obs. 2.

\nus\ consists of two independent focal plane modules FPMA and FPMB. The data were processed from both the modules using the standard \texttt{NuSTARDAS} 
software (version 1.8.0 of \texttt{HEASOFT} v.6.22.1 and CALDB version 20171002) to extract barycenter-corrected light curves, spectra, response matrices
and effective area files. The source events were extracted using a circular region of radius 49\arcsec\ and background events were extracted using a circle of same radius, away from the 
source.

The \swi/XRT data were analysed following standard procedures described in the 
\swi\ data analysis guide\footnote{\url{http://www.swift.ac.uk/analysis/xrt/}}. The source and background events were extracted
using circles of radii 45\arcsec. The task {\tt xrtpipeline} was used to generate the \swi/XRT spectrum. The response file was generated by using the task
\texttt{XRTMKARF}. 
 
\begin{figure}
\resizebox{0.8\hsize}{!}{\includegraphics[]{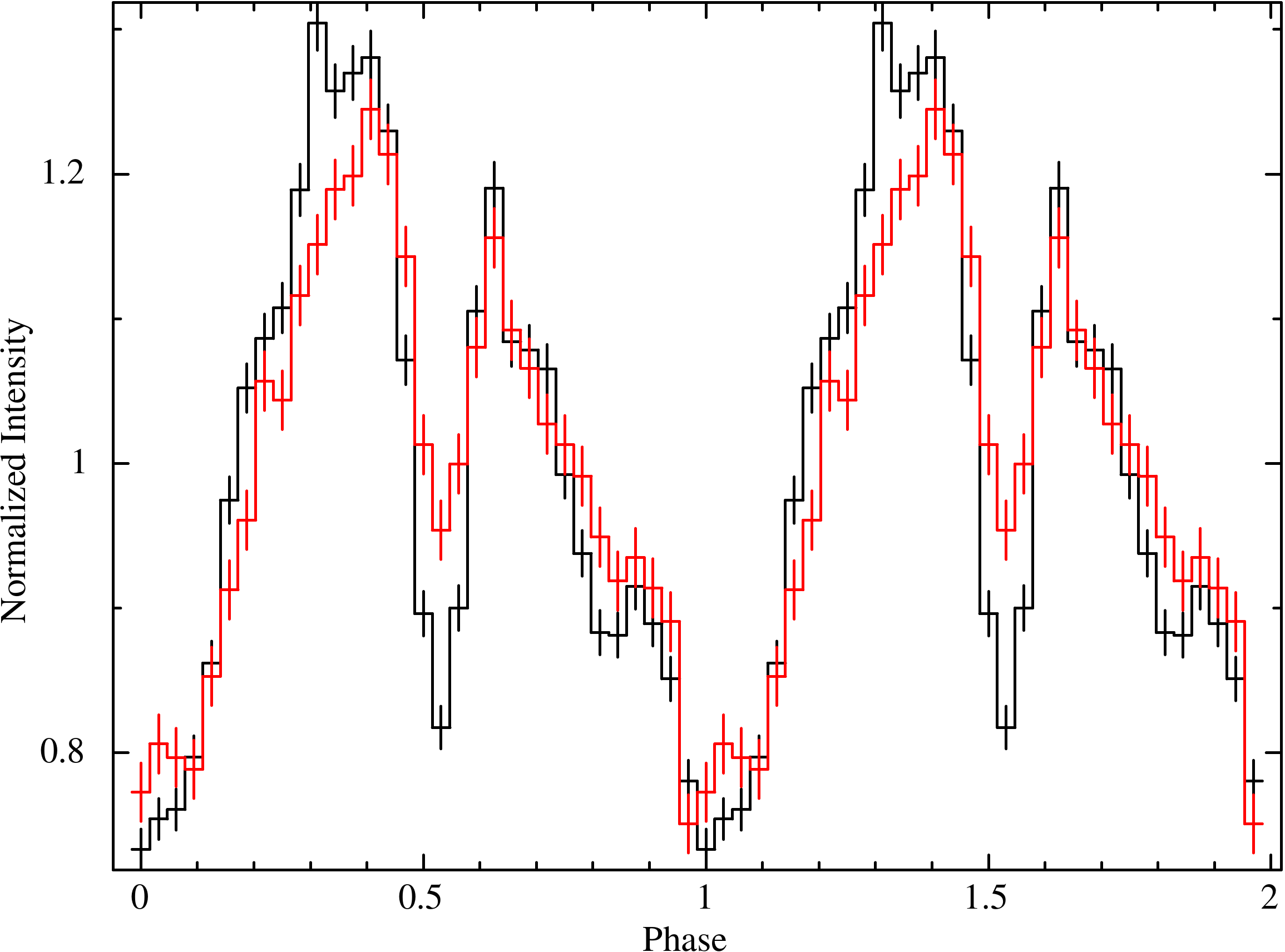}}
\caption{Background subtracted pulse profiles of \sxp, obtained from the FPMA detector of \nus\ (3--79\,keV, black)  and LAXPC10 of \astrosat\ (3--80\,keV, red).}
\label{pp}
\end{figure}
\begin{figure}
\resizebox{0.8\hsize}{!}{\includegraphics[]{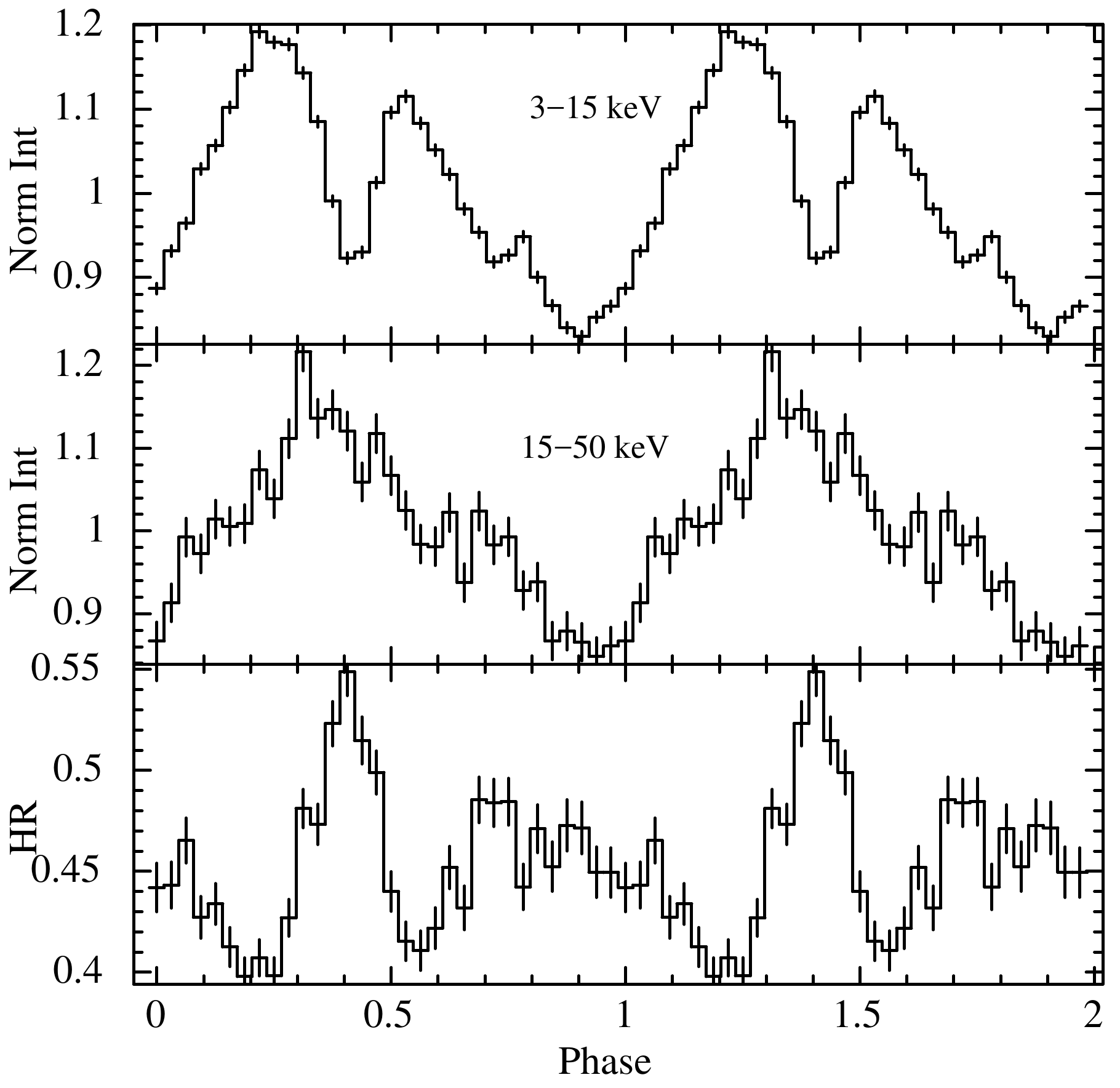}}
\caption{Background subtracted pulse profiles of \sxp\ obtained from LAXPC10 of \astrosat\ in the two energy bands of 3--15\,keV and 15--50\,keV and the HR variation with the pulse phase.
}
\label{hr}
\end{figure}

\astrosat\ consists of five independent instruments for performing simultaneous broadband observations. We analysed here data from the Soft X-ray Telescope (SXT) and the Large Area Xenon Proportional Counter (LAXPC).
SXT is a focusing telescope consisting of a CCD camera working in the energy range of 0.3--8 keV \citep{2014SPIE.9144E..1SS}.  
We used  level2 data (reprocessed from the level2 pipeline version 1.4a) and merged the event files using \texttt{sxtevtmergertool}. We extracted the spectrum thereafter using \texttt{XSELECT v2.4}\footnote{\url{http://www.tifr.res.in/~astrosat\_sxt/page1\_data\_analysis.php}} . The source events were extracted using an annular region between 1\arcmin and 16\arcmin, and an appropriate on-axis ARF was used for the spectral analysis. Blank sky SXT observations are used to extract the background spectrum. 

 The LAXPC consists of three co-aligned identical X-ray proportional counters having an absolute time resolution of 10 $\mu$s in the energy range 3.0--80.0 keV \citep{2017ApJS..231...10A,2017JApA...38...30A}.  
 Data were obtained in the  Event Analysis mode, level1 products of which were reprocessed with the LAXPC data analysis software\footnote{\url{http://www.rri.res.in/\~rripoc/}} to produce light curves 
 and spectral files. The events were filtered for the times  corresponding to Earth occultation, passage of the South Atlantic anomaly (SAA) and for large angle offsets of the detectors pointing away 
 from source. 
 The LAXPC background was taken from the same observation in periods when the source was occulted by Earth. LAXPC20 had a different gain during this observation, as determined from the k-fluorescent pulse amplitudes of
the double events. A gain factor was used to accommodate the same. All the three LAXPCs were used for the timing analysis. Due to a gas leakage in LAXPC30 leading to a loss of efficiency, LAXPC30 was excluded from the spectral analysis.
\section{Timing analysis}
\label{time}
We extracted light curves at 100\,ms time resolution for the timing analysis. 
 We used the pulse folding and $\chi^2$-maximisation method to estimate 
the barycentric corrected pulse period of the pulsar.  Pulsations were detected at 15.2563$\pm$0.0005 s in Obs. 1 and 15.2575$\pm$0.0009 s in Obs. 2 respectively. The errors correspond to 1$\sigma$ confidence. 
In the case of the LAXPC detectors (\astrosat), background subtraction was performed by subtracting the background count rates in the energy bands concerned. Fig. ~\ref{pp} shows the background subtracted pulse profiles from 
\astrosat\ and \nus. Pulsations are detected up to 50\,keV, and the pulse profiles from both the observations exhibit a double peaked profile which morphs to a single broad pulse at energies $>15$\,keV (Fig.~\ref{hr}). The pulse fraction was computed as the ratio of $(I_{max}-I_{min})/(I_{max}+I_{min})$. 
The pulsed fraction increases from 30\% in the energy range of 3--5\,keV to 60\% in the energy range of 30--50\,keV in Obs. 1, and from 20\%  in the energy range of 3--5\,keV to 40\% in 
the energy range of 30--50\,keV in Obs. 2. The disappearance of the double-peaked nature of the pulse profile with energy motivated us to investigate the hardness ratio (HR) along the spin phase 
of the system in two energy bands, 3--15\,keV and 15--50\,keV (Fig.~\ref{hr}). The HR was defined as the ratio of intensity in the 15--50\,keV band divided by the 3--15\,keV band. The HR shows significant evolution with the spin phase with a spectral hardening seen at the dip phase (phase $\sim$0.5).

\begin{figure}
   \resizebox{\hsize}{!}{\includegraphics[]{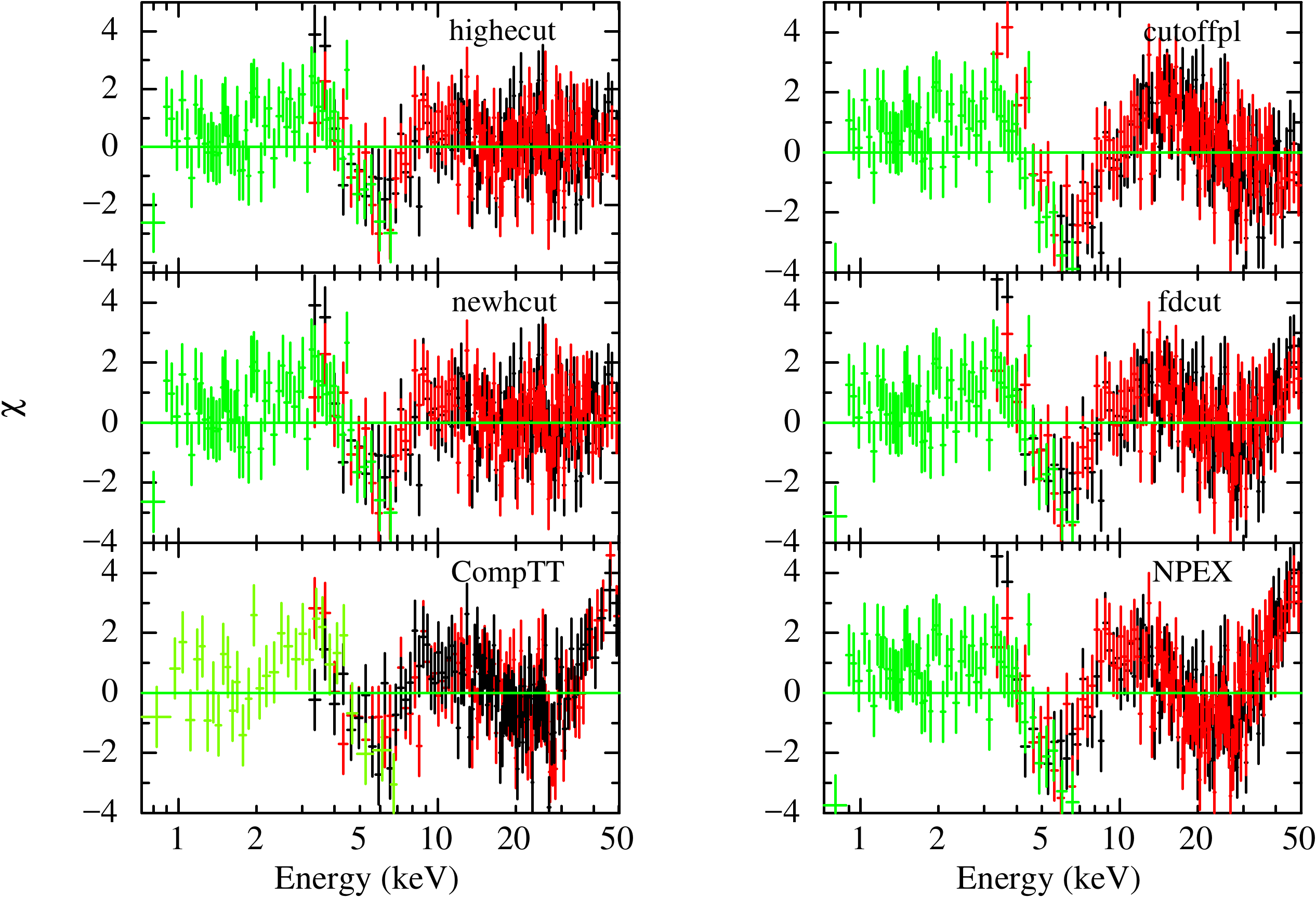}}
  \caption{The residuals of the spectral fits with \swi/XRT (in green), and FPMA ( in black) and FPMB (in red) detectors onboard \nus\ (Obs. 1). The continuum models are mentioned in the panels. An absorption feature at $\sim$5 keV is not included in the fits. }
  \label{crsf}
\end{figure}
\begin{figure}
  \resizebox{0.9\hsize}{!}{\includegraphics[]{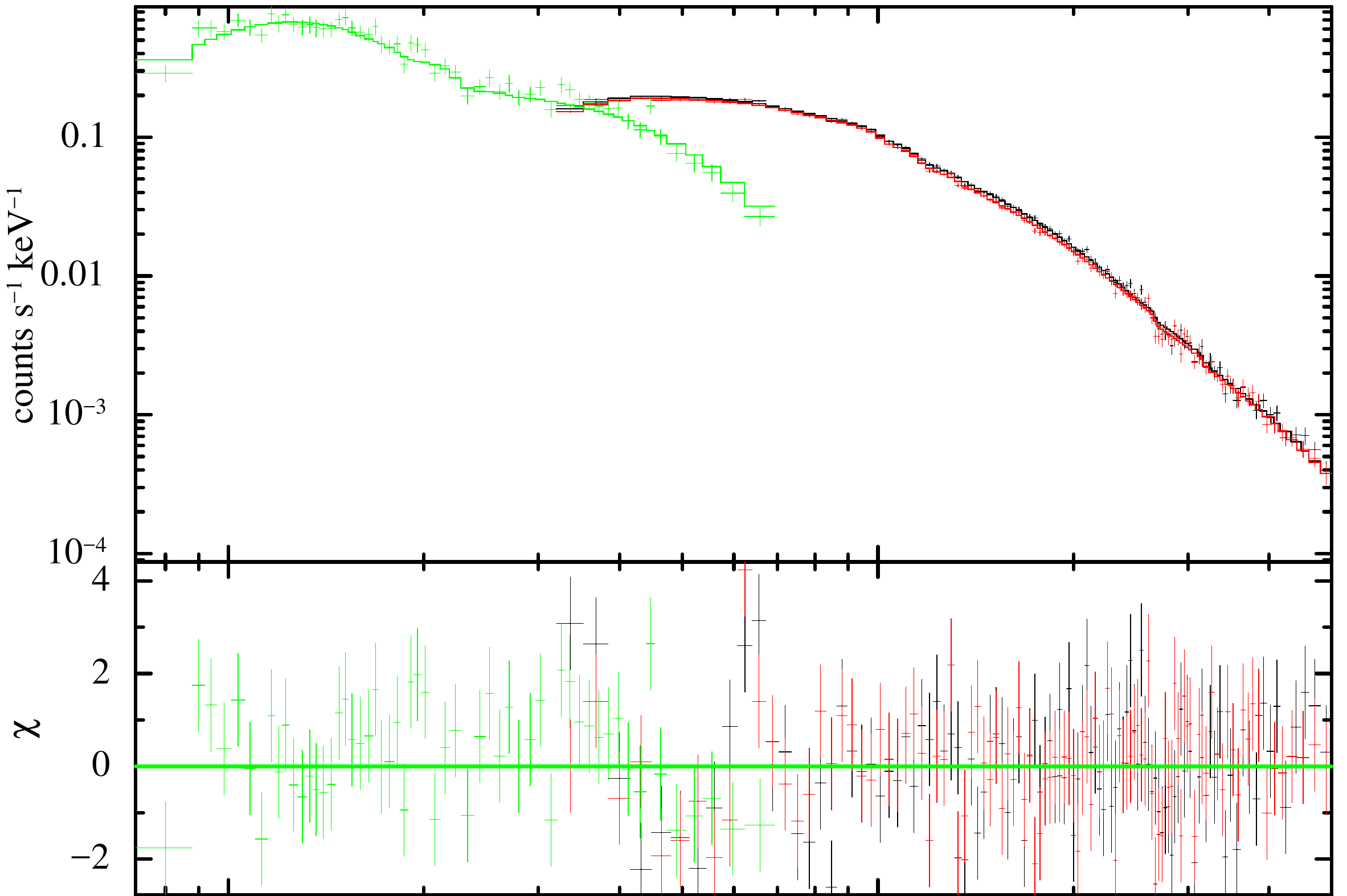}}
 \resizebox{0.902\hsize}{!}{\includegraphics[]{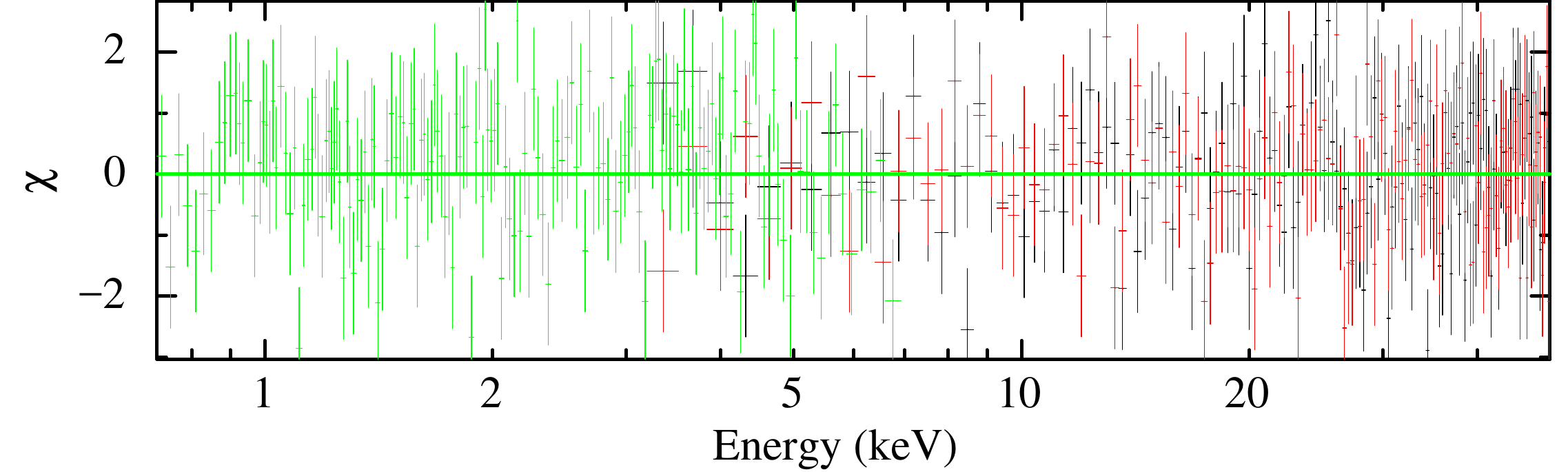}}
  \caption{The upper panel shows the best-fit spectral model of \sxp\ using spectra from  \swi/XRT (in green), and FPMA ( in black) and FPMB (in red) detectors onboard \nus\ (Obs. 1).  The second panel shows the residuals after the fit without taking into account the CRSF and the Fe line.  The third panels shows the residuals after including all the model components.}
  \label{spec}
\end{figure}
\begin{figure}
  \resizebox{0.9\hsize}{!}{\includegraphics[]{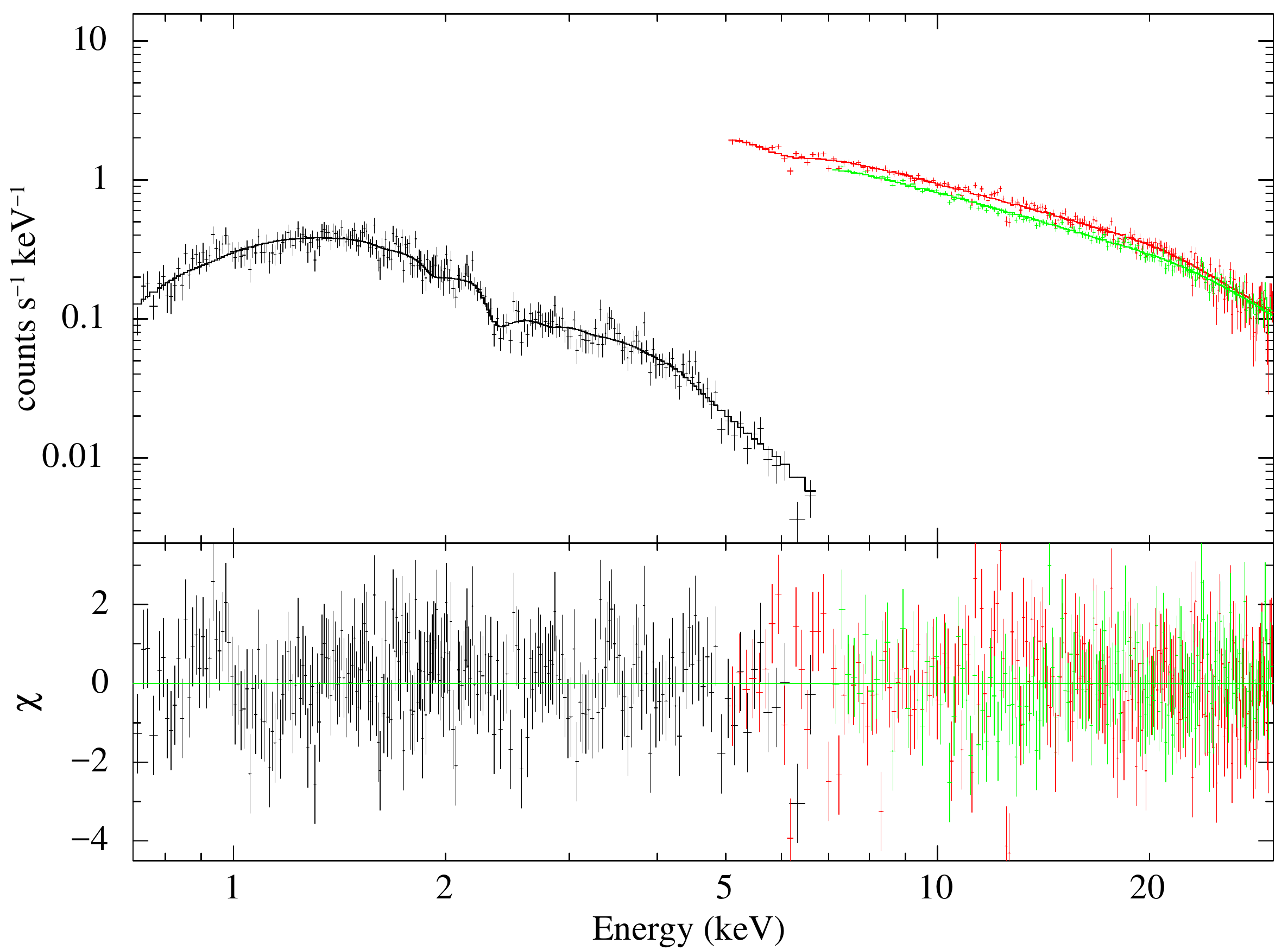}}
  \caption{Same as in Fig.~\ref{spec} using spectra from the SXT (in black), LAXPC10 (in red) and LAXPC20 (in green) onboard \astrosat\ (Obs. 2).
  The lower panel shows the residuals after including all the model components.}
  \label{spec2}
\end{figure}

\section{Broadband spectral analysis}
\label{sec:spec}
Spectral analysis was performed using \textit{XSPEC} v12.9.1. We grouped the spectra to achieve a minimum of 20 counts per spectral bin for the analysis.  
We investigated the broadband spectrum of \sxp\ using simultaneous  \swi/XRT and \nus\ data (Obs. 1) and SXT and LAXPC data (Obs. 2).  The spectra were modelled with standard continuum models like a power-law with quasi exponential high energy cutoff 
having various functional forms ({\it XSPEC} models `highecut', `bknpow', `fdcut' and `newhcut'). Other continuum models are a combination of two power laws with different photon
indices but a common cutoff energy value called the Negative and Positive power laws with Exponential model ({\it XSPEC} model 
'NPEX' ), and a thermal Comptonization model ({\it XSPEC} model 'CompTT').  In order to account for the photoelectric absorption by the interstellar gas, two components were used.
The first component was fixed to the Galactic value of $6\times10^{20}$ cm$^{-2}$ \citep{1990ARA&A..28..215D}. The second component was left free to account for the absorption within the SMC. For the latter
component, the metal abundances were fixed at 0.2 solar, as is typical in the SMC \citep{1992ApJ...384..508R}. The atomic cross-sections  were adapted from \cite{1996ApJ...465..487V}.  
The X-ray absorption was modelled using the {\it XSPEC} model `tbabs'. Finally, to account for inter-calibration uncertainties of the instruments and small flux variations of the source 
during not fully simultaneous observing intervals (Obs. 1) we introduce normalisation factors between instruments.

We obtained the best fit to the continuum with the `newhcut' model. This model is a modified version of `highecut' which
has a smoothed region around the cutoff energy. The smoothing function is a third order polynomial with continuous 
derivatives \citep{2000ApJ...530..429B}.
An iron fluorescence emission line at $\sim$6.4 keV was detected in the \nus\ spectrum. Additionally, 
a narrow absorption feature was visible at $\sim$5 keV in the broadband spectra of Obs. 1 and Obs. 2, irrespective of the continuum model used (Fig.~\ref{crsf}). Addition of a Gaussian absorption feature 
({\it XSPEC} model `Gabs') improved the fit significantly and the reduced $\chi^{2}$ after adding the absorption feature decreased to $\sim$1. Table\,1 summarises the best-fit broadband spectral parameters 
and Figs.~\ref{spec} and \ref{spec2} show the broadband spectra from Obs. 1 and Obs. 2, respectively. 
The continuum parameters are consistent between  Obs. 1 and Obs. 2. with an indication of spectral softening in Obs. 2. The absorption corrected broadband luminosity (0.5--50 keV) 
is $9.1\times10^{37}$  ergs s$^{-1}$ for Obs. 1 and $1.2\times10^{38}$  ergs s$^{-1}$ for Obs. 2 respectively.
 
An absorption feature detected in the energy spectrum of HMXB pulsars 
is reminiscent of a cyclotron resonance scattering feature (CRSF). 
 A careful modelling of the broadband continuum spectrum is essential to detect and model shallow features such as the CRSFs \citep[see for example][]{muller2013}.
 Although in this case the addition of the line was required for all the tested continuum models, the line width was narrowest and best constrained with the  `newhcut' model (Fig.~\ref{crsf}).
 The improvement in $\chi^{2}$ after adding the CRSF to the  `newhcut'  model was significant, with $\Delta\chi^{2}=75.39$ for 3 d.o.f. in the case of  Obs. 1, and  $\Delta\chi^{2}=155$ for 3 d.o.f. 
 in the case of  Obs. 2, respectively.  Although, the CRSF was detected more prominently in Obs. 2, we avoided further interpretations of the CRSF and its parameters with this observation as the 
 line lies at the edge of the energy bands for both the SXT and LAXPC detectors and needs to be treated with caution. However an independent detection of the line at the same energy and with an 
 independent instrument gives us confidence on the obtained results.
 
\section{Phase-resolved spectroscopy}
\begin{figure}
  \resizebox{0.95\hsize}{!}{\includegraphics[]{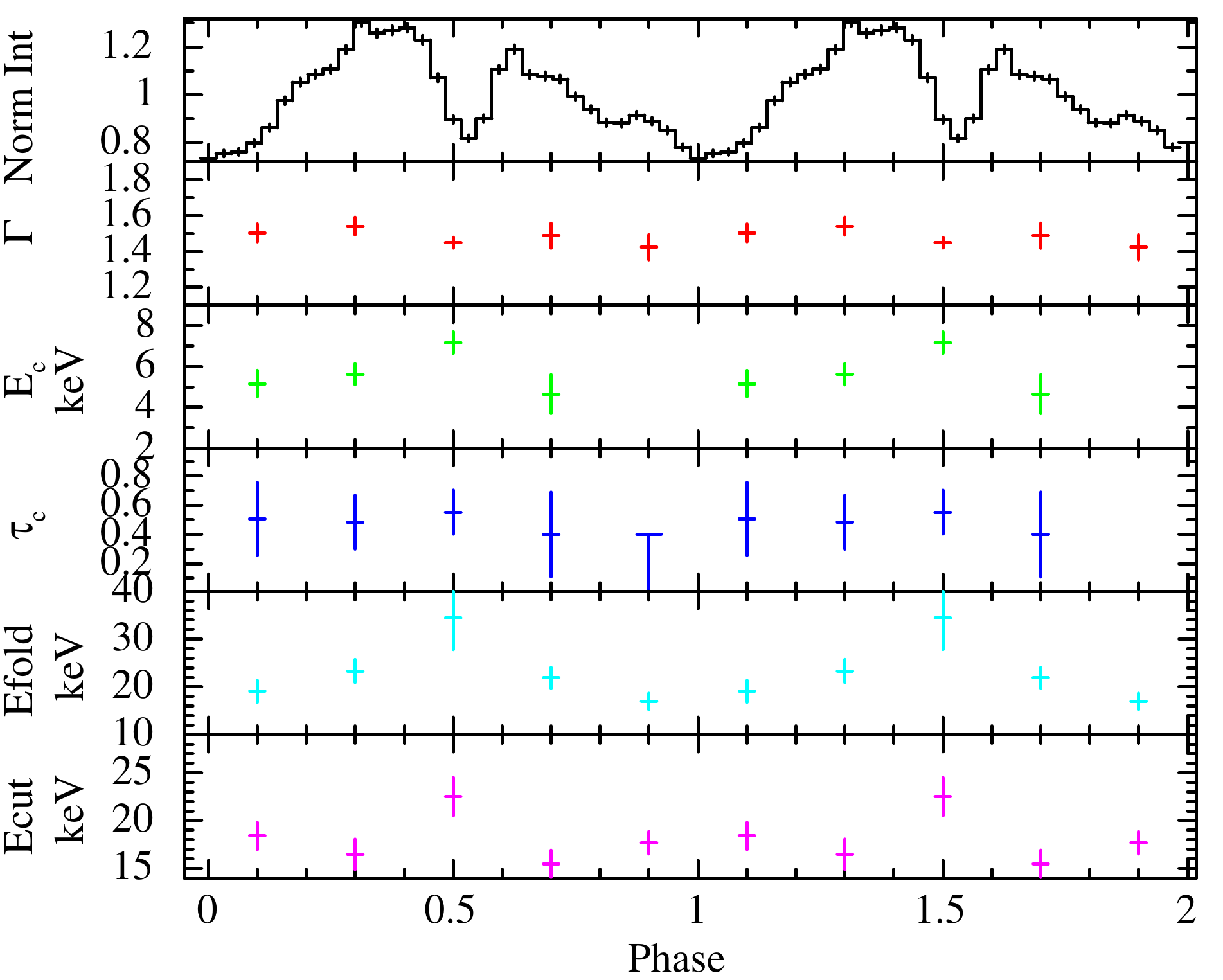}}
  \caption{Spectral parameters of \sxp\ obtained from the pulse phase-resolved spectroscopy from Obs. 1.
  The parameters are plotted with 90\% confidence. The top panel shows the pulse profile obtained from the FPMA detector (3--79\,keV).  }
  \label{phase}
\end{figure}

The variation of the HR with spin phase (Fig.~\ref{hr}) indicates a dependence of the spectral parameters on the changing viewing angle of the neutron star. 
Motivated by this we performed pulse phase-resolved spectroscopy using the \nus\ observation.
We created good-time-interval files (\texttt{gti}) using the measured pulse period of \nus\ to extract phase-resolved spectra into five equally
spaced phase bins. As the \swi/XRT data lacked the required statistics for the phase-resolved analysis of \sxp\ only \nus\ data were used for the purpose.  
The  `newhcut' continuum model was used for the spectral fits.  The SMC $\ensuremath{N_{\mathrm{H}}}$, the iron line energy and width, and the CRSF width ($\sigma_{c}$) 
were frozen to the phase averaged value in each phase bin. Fig.~\ref{phase} shows the variation of the spectral parameters with pulse phase.  
The spectrum is harder at the dip phase as compared to the peaks, i.e. $\Gamma=1.54\pm0.05$ at phase $\sim$0.3 to $\Gamma=1.40\pm0.03$ at phase $\sim$0.5. 
This is consistent with the results obtained from the investigation of the variation of the HR with the pulse phase. The CRSF centroid energy E$_{c}$ is variable 
with pulse phase with $E_{\rm c}$ rising to $\sim$8\,keV at the dip phase. $E_{\rm c}$ varies by a factor of $1.5\pm0.3$ between the dip and the adjacent phase 
bin (phase $\sim$0.7). The CRSF is not detected at the off-pulse phase which might be due to insufficient statistics in that phase bin. In order to obtain an upper 
limit on the CRSF depth at this phase, we froze the CRSF energy and width to the phase averaged value and obtained $\tau_{\rm c} \sim0.4$.  The variation of CRSF 
parameters with phase is typically seen in many HMXB pulsars, with the pattern of the variations revealing important information on the beaming geometry and the 
magnetic field geometry of the HMXB pulsar \citep[see][for a comprehensive summary of phase-resolved analysis of CRSFs]{2017JApA...38...50M}.

\begin{table*}
\centering
\caption{Best-fitting parameters (with 90\% errors) obtained from the spectral fitting 
with the newhcut continuum model with an iron emission line and cyclotron absorption line.}

\begin{tabular}{lcccc}
\hline
Parameter                       &\multicolumn{2}{c}{Obs. 1}       &\multicolumn{2}{c}{Obs. 2}       \\
                           &NEWHCUT             &NEWHCUT$\times$GABS   &NEWHCUT       &NEWHCUT$\times$GABS  \\
\hline

SMC $\ensuremath{N_{\mathrm{H}}}$$^a$  &0.40$\pm$0.09    &0.6$\pm$0.1  &0.31$\pm$0.06  &0.45$\pm$0.06  \\
Photon index               &1.40$\pm$0.01  &1.48$_{-0.03}^{+0.06}$   &1.40$\pm$0.02         &1.40$\pm$0.02 \\
E$_{\rm cut}$ (keV)	              &16.4$\pm$0.6      &16.9$\pm$0.8  &18.8$\pm$2.0            &20.9$\pm$2.3    \\
E$_{\rm fold}$ (keV)	  & 26.8$\pm$1.4 & 28.5$_{-1.7}^{+2.2}$ &37.1$_{-10.4}^{+12.8}$ & 33.4$_{-12.8}^{+17.6}$ \\
Fe line energy (keV)             &6.39$\pm$0.08  &6.37$\pm$0.08   & --         & --   \\
Fe line eq. width (eV)             &89.7$\pm$18      &91.4$\pm$18    & --              & --   \\
\\
Cycl. line energy (E$_{\rm c}$) (keV)           &--             &5.7$_{-0.6}^{+0.3}$    &--               &5.2$\pm$0.2     \\
Cycl. line width ($\sigma_{\rm c}$) (keV)         &--             &1.7$_{-0.5}^{+0.8}$    &--               &0.67$\pm$0.17    \\
Cycl. line strength ($\tau_{\rm c}$)	                &--        &  0.4$_{-0.2}^{+0.5}$  &--            &0.6$\pm$0.1      \\
\\
Luminosity$^b$   &     --             &0.91$\pm$0.05    & --            &1.2$\pm$0.1       \\ 
Reduced-$\chi^2$ (d.o.f)         &1.19 (450)     &1.04 (447)  &1.39 (582)          &1.13 (579)      \\
\hline
\end{tabular}
\\
\flushleft
$^a$ : Equivalent hydrogen column density (in 10$^{22}$ atoms cm$^{-2}$); 
$^b$ : Absorption corrected luminosity (0.5--50 keV) in 10$^{38}$  ergs s$^{-1}$, assuming a distance of 61~kpc. 
\\
\label{spec_par}
\end{table*}

\section{Discussion and conclusions}

In this letter we report the broadband X-ray timing and spectral properties of \sxp\ for the first time, and at the brightest state of the source detected till today. 
We also report the discovery of a CRSF at $\sim$5 keV. This makes it only the second Magellanic pulsar after SMC X-2 \citep{2016MNRAS.461L..97J} with a cyclotron line detection, 
and hence a confirmed magnetic field strength of the neutron star. The spin period measurements with the \nus\ and \astrosat\  observations are consistent within errors precluding the detection of any spin-up during the current outburst. The net spin-up rate between the {\it CGRO} and {\it ROSAT} observations separated by 123 days was  $-1.64\times10^{-9}$\,s\,s$^{-1}$ \citep{2001ApJ...560..378F}. The long term trend in the spin evolution as inferred from the spin period measurements between the {\it CGRO} and \astrosat\  observations however indicate a much reduced spin-up rate of  $-2.92\times10^{-11}$\,s\,s$^{-1}$.

The magnetic field strength of the neutron star can be determined from the observed CRSF centroid energy 
$E_{\mathrm{cyc}}$ (determined from Obs. 1), and is given as:
\begin{equation}
	E_{\mathrm{cyc}} = \frac{11.57\,\mathrm{keV}}{1+z} \times B_{12}
	\label{eqn:12b12}
\end{equation}
where $B_{12}$ is the field strength in units of $10^{12}$\,G; $z \sim 0.3$ is the gravitational red shift in the scattering region for standard neutron star parameters. This implies a magnetic field strength of the neutron star of $B=6\times10^{11}$ G, assuming the line forming region lies close to the neutron star surface. The obtained field strength is consistent with the estimate
 obtained by \cite{2017RAA....17...59C} assuming that \sxp\ was in the propeller state at its lowest detected X-ray luminosity ($L_{\rm x} \sim6.8\times10^{33}$ erg s$^{-1}$ as detected from a \cha\ observation). 

The unabsorbed bolometric X-ray luminosity of \sxp\ during the observations indicate that the source was accreting near its Eddington limit of $2\times10^{38}$ erg s$^{-1}$ for a typical neutron star mass of 1.4 $\mathrm M_{\odot}$.  In highly magnetised accretion powered pulsars, the location and geometry of the radiation emitting region are believed to be dependent on the mass accretion rate \citep{1976MNRAS.175..395B}.  
At a luminosity of $\sim$10$^{38}$ erg s$^{-1}$  \sxp\ is expected to be in the super-critically accreting or radiation-dominated regime. In the  super-critical regime, a radiation-dominated shock is formed, after which the accreted matter settles to the neutron star
 surface in a magnetically confined accretion column. The radiation in this case predominantly escapes from the optically thin sides of the accretion column in a fan-beam like pattern. 
The critical-luminosity ($L_{\rm c}$), which divides the two regimes of sub and super-critical accretion is a function of the surface magnetic field strength of the neutron star and can be approximated as \citep{2012A&A...544A.123B}:

\begin{eqnarray}
	L_{\rm c} &=& 1.49 \times 10^{37}{\rm erg\,s}^{-1} \left( \frac{\Lambda}{0.1} \right)^{-7/5} w^{-28/15} \nonumber \\
	&&\times \left( \frac{M}{1.4{\rm\,M}_{\odot}} \right)^{29/30} \left( \frac{R}{10{\rm\,km}} \right)^{1/10} \left( \frac{B_{\rm surf}}{10^{12}{\rm\,G}} \right)^{16/15}
	\label{eqn:lcrit}
\end{eqnarray}
$M$, $R$, and $B_{\rm surf}$ are, the mass, radius, and surface
magnetic field strength of the neutron star, $w = 1$ characterises the shape of the
photon spectrum inside the column, and $\Lambda$ is the mode of
accretion. $\Lambda
= 0.1$ approximates the case of disk accretion, and $\Lambda
= 1.0$ is more appropriate for wind accretors. Assuming  $\Lambda
= 0.1$ and $B_{\rm surf}=6\times10^{11}$ G, results in $L_{\rm c}=9\times10^{36}
$ erg s$^{-1}$ for typical neutron star mass and radius values\footnote{A more detailed treatment of critical-luminosity can be found in \cite{2015MNRAS.447.1847M}. 
We however verified that our obtained $L_{\rm c}$ is consistent between the two works for the estimated $B_{12}$ value.}. This ascertains that \sxp\ is accreting in the super-critical regime.
The double-peaked pulse profile of \sxp\ observed in this work is in further support of the predominance of a fan-beam like emission. The disappearance of the double peak to a single broad 
peak at higher energies is most likely due to the intrinsic nature of the emission rather than being caused by a local absorbing matter phase locked to the neutron star. This is because we 
found no evidence of an additional absorption component in the spectral fit of \sxp.
 A further indication of the fan-beam emission is obtained from the shape of the CRSF. A deep and narrow CRSF, as seen in \sxp\ is expected for viewing angles perpendicular to the magnetic field axis, a.k.a. fan-beam like emissions \citep{2017A&A...601A..99S}. 
 
 The luminosity of \sxp\ during Obs. 2 was $\sim$30\% higher than in Obs. 1, with an indication of spectral softening with increasing luminosity. This behaviour is expected in the  
 super-critical regime, and can be understood either due to a decrease in the plasma temperature with rising accretion column \citep{2012A&A...544A.123B}, or alternatively a lower fraction of the radiation reflected by the neutron star surface in the case of a taller accretion column at higher intensities \citep{2015MNRAS.452.1601P}.
 
The CRSF parameters show little variations with pulse phase. This may indicate no gradient of the properties of the line forming region across the viewing angles. 
Alternatively, this might also be due to the effect of gravitational light bending near the neutron star surface \citep{2002ApJ...566L..85B} which would smear out the pulse-phase dependence, 
with a particular viewing angle having contributions from multiple emission regions. The only variable CRSF parameter is the centroid energy E$_{c}$ which is significantly higher at the 
dip phase.  This suggests that the line forming region at this phase may offer a deep and a more direct view into the emission region along the magnetic axis which is consistent with fan-beam 
like emission.  An indication of spectral hardening at the dip phase is further consistent with a direct view into the emission region along the magnetic axis \citep[see for e.g.][]{1978ApJ...225..988P}.

In summary we present the broadband timing and spectral results of \sxp\ for the first time using \astrosat\ and \nus\ ToO observations performed
during the recent outburst in October 2017. We also report the discovery of a CRSF at $\sim$5 keV, establishing the magnetic field of the neutron star at $6\times10^{11}$ G. 
The CRSF centroid energy varies with pulse phase, with an increase in energy during an intensity dip. This is accompanied with a spectral hardening during the dip. The two 
signatures mentioned above and the double-peaked pulse profile of \sxp\ indicate a fan-beam like geometry dominating the emitting region as is expected for 
super-critically accreting sources.


\section*{Acknowledgements}
We thank the reviewer for providing useful comments. The authors would like to thank  the AstroSat operations team members for scheduling the ToO observation.
The authors would also like to thank all the \nus\ and \swi\ team members for ToO observations.
CM acknowledges Gulab Dewangan for useful discussions and suggestions on the analysis of the SXT data.



\bibliographystyle{mnras}
 \bibliography{article1}
 \label{lastpage}
 \end{document}